\documentclass[12pt]{article}

\usepackage{array}
\usepackage{epsfig}
\usepackage{amssymb}
\usepackage{graphics,graphpap}
\usepackage{multicol}
\usepackage{float}
\usepackage{slashed}
\usepackage{multirow}
\usepackage{amsmath}
\usepackage{tabularx}

\RequirePackage[sort&compress,square,comma,numbers]{natbib}
\renewcommand{\thefootnote}{\fnsymbol{footnote}}

\def\calAslash{\rlap{\hspace{0.08cm}/}{{\cal A}}}

\newcommand{\state}[1]{|#1\rangle}

\begin{document}

\begin{titlepage}
\vskip1.5cm
\begin{center}
   {\Large \bf Study on pure annihilation type \boldmath $B \to V\gamma$ decays}
    \vskip1.3cm
    { Hui Deng$^a$, Jing Gao$^{b,c}$, Lei-Yi Li$^{d,b}$\footnote{Corresponding author: leiyi@mail.nankai.edu.cn}, Cai-Dian L\"u$^{b,c}$, Yue-Long Shen$^a$\footnote{Corresponding author: shenylmeteor@ouc.edu.cn}, Chun-Xu Yu$^d$}\\
    \vskip0.5cm
        { \it $^a$ College of Information Science and Engineering, Ocean
         University of China, Qingdao 266100,  China }\\
        { \it $^b$ Institute of High Energy Physics, CAS, P.O. Box 918(4) Beijing 100049,  China} \\
        { \it $^c$ School of Physics, University of Chinese Academy of Sciences, Beijing 100049, China} \\
        { \it $^d$ School of Physics, Nankai University, Weijin Road 94, Tianjin 300071,  China} \\
\vskip2.5cm
{\large\bf Abstract\\[10pt]}
 \parbox[t]{\textwidth}{

We investigate   the pure annihilation type radiative $B$ meson decays $B^0 \to \phi \gamma$ and $B_s \to \rho^0(\omega)\gamma$ in the soft-collinear effective theory. We consider three types of contributions to the decay amplitudes, including the direct annihilation topology, the contribution from the electro-magnetic penguin operator and  the contribution of the neutral vector meson  mixings.   The numerical analysis shows  that the decay amplitudes are dominated by the $\omega-\phi$ mixing effect in the $B^0 \to \phi\gamma$ and $B_s \to \omega\gamma$ modes.  The corresponding decay branching ratios are enhanced about three orders of magnitudes relative to the pure annihilation type contribution in these two decay channels. The decay rate of $B_s \to \rho^0\gamma$ is much smaller than that of $B_s \to \omega\gamma$ because of  the smaller   $\rho^0-\phi$ mixing. The  predicted branching ratios $B(B^{0}\rightarrow\phi\gamma)=(3.99^{+1.67}_{-1.46} )\times10^{-9},\,B(B_s\rightarrow\omega\gamma)=(2.01^{+0.81}_{-0.71} )\times10^{-7}$
 are to be tested by the Belle-II and LHC-b experiments.
}

\end{center}

\vfill

\end{titlepage}

\setcounter{footnote}{0}
\renewcommand{\thefootnote}{\arabic{footnote}}

\newpage

\section{Introduction}\label{sec:1}

The exclusive radiative $B$ decay modes $B \to V\gamma $ are very
interesting and valuable probes of flavor physics, since they
provide an excellent platform to constrain standard model
parameters, to test new physics models and to understand QCD
factorization of the decay amplitudes\cite{Hurth}. Most $B \to
V\gamma $ decays occur via the flavour-changing neutral-current
 transitions $b\to s\gamma$ or $b\to d\gamma$,   and the
quark level transition amplitudes are now approaching
next-to-next-to-leading order accuracy \cite{misiak,NNLObsgamma}.
It is more  profound to evaluate the exclusive decay modes $B \to
V\gamma $, based on the effective theory with the
expansion of the inverse powers of the $b$ quark mass. At leading power
in $1/m_b$, the QCD factorization  of $B \to
V\gamma $ decays has been established up to next-to-leading order in $\alpha_s$
\cite{Beneke:2001at,Beneke:2004dp,BVga1,AP,BoBu1,BoschThesis,BoBu2,BZ06b}. The leading power factorization formula was confirmed in a more elegant way with soft-collinear effective theory (SCET) \cite{BHN}.
The exclusive $B\to V\gamma$
decays have also been investigated in the alternative approach of
perturbative QCD factorization  based on $k_T$ factorization \cite{pQCD}.

In the modern accelerators with high luminosity,  more accurate data have been accumulated,  therefore, besides the leading power contributions we must consider power corrections on the theoretical side to improve the theoretical  precision.
  Among the power suppressed corrections, the weak annihilation
diagrams are of great importance as it might be mediated by tree
operators, and they play an important role on the determination of the
time-dependent CP asymmetry in $B\to V\gamma$, see
Refs.~\cite{alt,grin04,grin05,cpas}, as well as isospin
asymmetries \cite{kagan}. There exists a special type of
radiative decays that the decay amplitude contains only annihilation type diagrams,
including $B^0 \to \phi\gamma$ and $B_s \to \rho^0(\omega) \gamma$
decays. Relatively less attentions are paid to them due to their
tiny branching ratios \cite{Li:2003kz}. The $B^0 \to
\phi\gamma$ decay is mediated by penguin annihilation topology,
with very small Wilson coefficient. In addition, this decay mode
is suppressed by $\Lambda/m_b$, since the emitted vector meson must be
transversely polarized. In naive factorization, its branching
ratio is estimated to be at the order of $10^{-13}$, and QCD
corrections can enhance the result to about $10^{-12}$.  In ref.\cite{Lu:2006nza}, it was found that the electromagnetic penguin
operator $O_{7\gamma}$ contribution through $B^0 \to \gamma\gamma^\ast $
with the virtual photon connecting to the $\phi$ meson can increase the
branching ratio for $B^0 \to \phi\gamma$ to the order of $10^{-11}$. The predicted branching ratios within the framework of perturbative QCD factorization approach is   also at this order \cite{pureann}.
For $B_s \to \rho^0(\omega) \gamma$ mode, the contribution from
electromagnetic penguin operator  is also of great importance, and the
branching ratio is at the order of $10^{-10}\sim10^{-9}$\cite{Lu:2006nza}.

On the other hand, the radiative decays mediated  by tensor transition form factors
 have much larger branching ratios. The measured branching fractions of $B^0 \to
\rho^0(\omega)\gamma$ and $B_s \to \phi \gamma$ read \cite{Tanabashi:2018oca}
\begin{eqnarray}
B(B^0 \to \rho^0 \gamma)&=&(8.6\pm 1.5) \times 10^{-7},\nonumber \\
B(B^0 \to \omega \gamma)&=&(4.4^{+1.8}_{-1.6})\,\,\, \times 10^{-7},\nonumber \\
B(B_s \to \phi \gamma)&=&(3.4\pm 0.4) \times 10^{-5}.
\end{eqnarray}
They are at least four orders larger than the predicted $B^0\to \phi
\gamma$ and $B_s \to \rho^0(\omega)\gamma$ decays. Such a large
discrepancy might leads to large contribution to $B^0 \to
\phi\gamma$ and $B_s \to \rho^0(\omega) \gamma$ decays through the mixing
between the neutral vector mesons $\omega, \rho^0$ and $\phi$. The
$\omega-\phi$ mixing effect is regarded to be large in many $B$
meson and $D$ meson decays modes \cite{Gronau:2008kk,Gronau:2009mp,Li:2009zj}. Thus it is
valuable to investigate n the contribution of this effect in purely
annihilation type $B \to V\gamma $ decays, which may be the dominant contribution.
If the branching ratio of pure annihilation type $B$ decays can be significantly enhanced by the
neutral meson mixing, the super-B factory and LHC-b  might have the chance to find the signals of these
processes.

This paper is arranged as follows: In the next section we will
present the factorization formulas of $B^0 \to \phi \gamma$ and $B_s
\to \rho^0(\omega) \gamma$ decays, including the leading power contribution,  and the contributions from the annihilation topology, the
electro-magnetic penguin operator and the $\omega-\phi$ mixing effect. Numerical
analysis will be presented in Section 3.  The last section is
closing remarks.

\section{Theoretical overview of pure annihilation type radiative $B$($B_s$) decays}\label{sec:2}

The effective Hamiltonian for $b\to D\gamma$ transitions, with
$D=s,d$, reads:
\begin{eqnarray} \label{heff} {\cal H}_{\rm eff} = \frac
{ G_F}{\sqrt{2}}\sum\limits_{p=u,c}\lambda_p \left[C_1
O^p_1+C_2 O^p_2+\sum_{i=3}^{10} C_i O_i+C_{7\gamma}O_{7\gamma}+C_{8g}O_{8g}
\right ]+h.c. \,, \end{eqnarray}
where  $\lambda_p=V^*_{pD} V_{pb}$ and $V_{ij}$ are
elements of the CKM matrix, $O^{(p)}_i(\mu)$ are the relevant
operators and $C_i(\mu)$ are the corresponding Wilson
coefficients, which are shown in ref.\cite{beyondLL,buras,Beneke:2000ry}.

The $B \to V \gamma$ decays contain several kinds of momentum
modes, which is convenient to work
in the light-cone coordinate system, where the collinear momentum
of the vector meson $p$ can be expressed as
$p=(n \cdot p,{\bar n}\cdot p,p_\perp) \sim (\lambda^2, 1,\lambda)m_b~$, with
the null vector $n$ and $\bar n$ satisfying $n\cdot \bar
n=2$. The anti-collinear photon momentum $q$ scales as $ ( 1,\lambda^2,
\lambda)m_b$. In addition, the momentum of soft quark inside the $B$ meson
 and intermediate hard-collinear quark or gluon can be expressed as $(\lambda,\lambda,\lambda)m_b$ and $(\lambda,1,\sqrt{\lambda})m_b$ respectively. All these modes are  necessary
to correctly reproduce the infrared behavior of full QCD. SCET provides a more transparent language of the factorization of multi-scale problems than diagrammatic approach. In SCET, the fields with typical momentum mode have definite power counting rules. The power behaviors of the fields appear at $B \to V\gamma$ decays
are as follows
\begin{eqnarray}
&\xi_c \sim \lambda\,,\,\,\,\,\,
A_{c}^\mu \sim (\lambda^2, 1,
\lambda)\,,\,\,\,\,\,\xi_{hc} \sim \lambda^{1/2}\,,\,\,\,\,\,
A_{hc}^\mu \sim (\lambda, 1,
\lambda^{1/2})\,, \nonumber \\
&\xi_{\bar c} \sim \lambda\,,\,\,\,\,\, A_{\bar c}^\mu \sim ( 1,
\lambda^2,\lambda)\,,\,\,\,\,\,   q_s \sim \lambda^{3/2}\,,\,\,\,\,\, A_{s}^\mu \sim (\lambda,
\lambda, \lambda)\,,\,\,\,\,\, h_v \sim \lambda^{3/2}\,.
\end{eqnarray}
 Since SCET contains two kinds of collinear fields, {\it i.e.}
hard-collinear and collinear fields,  an intermediate
effective theory, called $\rm SCET_I$, is introduced that
contains  soft, collinear and hard-collinear fields.   The
final effective theory, called $\rm SCET_{II}$, contains only soft
and collinear fields. To obtain the amplitudes of  radiative
decays, one needs to do a two-step matching from ${\rm QCD} \to {\rm
SCET_I} \to {\rm SCET_{II}}$. The matching procedure of leading power amplitude
has been performed in ref.\cite{BHN}.   In the following we give a brief review in order that we can conveniently express the contribution of the neutral meson mixing.

In the first step the hard scale $m_b^2$ is integrated
out by matching the operators $Q_i$ in the weak Hamiltonian onto
a set of operators in ${\rm SCET_I}$.
Merely considering the operators contributing at leading power, the matching  takes the form
\begin{equation}
 {\cal H}_{\rm eff} \to
 C^{A}  Q^{A} +  C^{B}\otimes Q^{B}.
\end{equation}
The $\otimes$ denotes a convolution over space-time or momentum fractions.  The
momentum-space Wilson coefficients depend only on quantities at
the hard scale $m_b^2$. The specific form of the operators $Q^{(i)}$
is written by:
\begin{eqnarray}\
\label{eq:SCETJs}
Q^{A}&=&\left(\bar\xi W_{hc}\right)(s\bar n)\not\!\! {\cal A}^{em}_{c\perp}(tn)(1-\gamma_5) h_v, \nonumber \\
Q^{B}&=&\left(\bar\xi W_{hc}\right)(s\bar n)\not\!\! {\cal A}^{em}_{c\perp}(tn)
\calAslash_{hc_\perp}(r\bar n)
(1+\gamma_5)  h_v .
\end{eqnarray}
 The definition of SCET$_{\rm I}$ building block ${\cal A}_{hc}$ and Wilson line $W_{hc}$ have been given in \cite{BHN}.  The $B$-type operators are actually power
suppressed in ${\rm SCET_I}$, but contribute at the same order as
the $A$-type operator upon the transition to ${\rm SCET_{II}}$.

The matrix element of the operator $Q^{A}$ is proportional to the
SCET form factor $\zeta_{V_\perp}$, i.e.,
\begin{eqnarray}
\langle V_\perp(\varepsilon_1)|\bar \xi \Gamma h_v| \bar
B_v\rangle=2E\zeta_{V\perp}(E){\rm Tr}\left[{\not\! n\not\!\bar
n\over 4}\not\!\varepsilon_{1\perp}^*\Gamma{1+\not\! v\over
2}\gamma_5\right].\end{eqnarray}
 The operators $Q^{B}$ can be
further matched onto four-quark operators in ${\rm SCET_{II}}$ through
time ordered product with ${\rm SCET_I}$ Lagrangian \cite{Beneke:2002ph}
\begin{eqnarray}
\int d^4x \langle V_\perp \gamma|T\{{\cal L}_{\xi q}^{(1)}(x),Q^{B}(0)\}| \bar
B_v\rangle=\int ds\int
dt \tilde J_\perp(s,t) O^{B}(s,t).\end{eqnarray} with
\begin{eqnarray}
 O^{B}(s,t)=[\bar\xi W_c](s\bar n)(1+\gamma_5)\not\!\!{\cal A}_{\perp\bar{
 \rm c}}^{\rm em}{\not\!\bar n\over 2}[W_c^\dagger\xi](0)
 [\bar q_sY_s](tn)(1-\gamma_5){\not\! n\over 2}h_v(0).\end{eqnarray}
 When matching the
operator $Q^{B}$ onto ${\rm SCET_{II}}$, the hard-collinear scale
$m_b \Lambda$ is integrated out, and the matching
coefficient gives rise to the jet function
\begin{equation}
\, J^B_\perp(\omega,u) = \int dt \, e^{-i\omega t}\int ds
e^{-2iEus} \, \tilde J^B_\perp(s,t)  \,.
\end{equation}
The final low-energy theory ${\rm SCET_{II}}$ contains only soft and
collinear fields. At leading power the factorization theorem is proved in an elegant way with
SCET since the soft and collinear fields decouple.  The
soft fields are restricted to the $B$-meson light-cone distribution amplitude (LCDA), and collinear
ones to the vector meson LCDA defined as
\begin{eqnarray}
\langle 0| [\bar{{ q}} _{s}Y_s](tn)\,{\not\! n\over 2}\Gamma\,
Y_s^\dagger{h}_v(0)
 |\bar B_v \rangle
&=& - \frac{i F(\mu)\sqrt m_{B}}{2}\,{\rm tr}\bigg[{\not\! n\over
2}
 \Gamma \,\frac{1+\!\not\!v}{2}  \gamma_5 \bigg]
\int_0^\infty d\omega\,e^{-i \omega t n\cdot
v}\,\phi^+_B(\omega,\mu) \, \nonumber \\
\langle V(p)| [\bar{\xi}W_c](s\bar n)\,\Gamma\,{\not\! \bar n\over
2} W_c^\dagger {\xi}(0)
 | 0 \rangle
&=&  \frac{i f_V\bar n\cdot p}{4}\,{\rm tr}\bigg[
 {\not\!  n\not\!\bar  n\over 4}\not\!\varepsilon^*\Gamma  {\not\!\bar  n\over 2} \bigg]
\int_0^1 d\omega\,e^{i us\bar n\cdot p}\,\phi_V(u,\mu) .
\end{eqnarray}
 The final
factorization formula is then written by
\begin{eqnarray}
\label{eq:SCETffQCDff} \left \langle V \gamma \left | {\cal
H}_{\rm eff} \right | \bar B \right \rangle|_{\rm LP} &=&2m_B\left
[ C^A \zeta_{V_\perp}+\frac{\sqrt{m_B}F(\mu) f_{V_\perp}}{4}
\left(C^{B}\otimes J_\perp\right) \otimes \phi^V_\perp
\otimes\phi^B_+\right] .
\end{eqnarray}
Up to the order  of $ \alpha_s$, the explicit expression of hard functions $C^A(\mu)$ and  $C^B(\mu)$
have been given in \cite{BHN}(here we use $C^B$ instead of $C_1^B(\mu)$). The leading power contribution is dominant in the decays $B^0 \to \rho^0(\omega)\gamma$ and $B_s \to \phi \gamma$, which will be employed in our evaluation of the contribution from mixing of neutral vector mesons.

\begin{figure}
\begin{center}
\hspace{0cm}\scalebox{1.2}[1.2]{\includegraphics[width=0.5\textwidth]{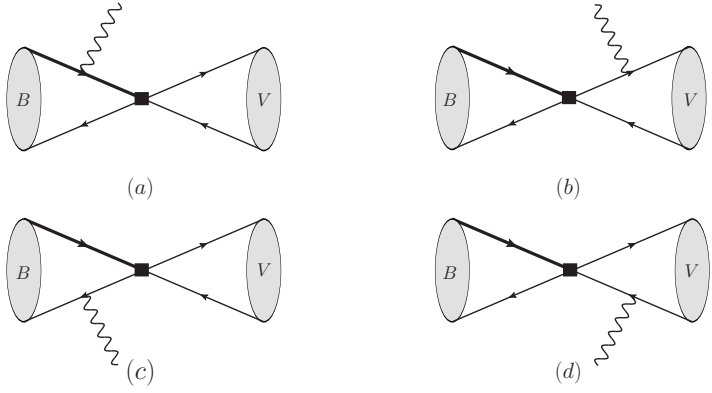}}
\caption{\label{figwa}Weak annihilation diagrams for $B^0 \to \phi \gamma$ decay.
}\end{center}
\end{figure}

\subsection{Contribution from weak annihilation}

Now we are ready to investigate the weak annihilation   contribution
to the purely annihilation type operators. The  weak annihilation diagrams are shown in Fig.~\ref{figwa}, all of which are
mediated by the four-quark operators.   In order to produce a
transversely polarized vector meson, the four-quark operators must
be matched to the $\rm SCET_I$ operators which are suppressed by
$1/m_b$. Among the four diagrams in Fig.~\ref{figwa}, diagram (c) is dominant because it is enhanced by
a hard-collinear propagator compared with the other diagrams. Therefore we neglect diagrams (a,b,d), which are highly suppressed in our calculation.
 Only considering the contribution of the leading two particle Fock
state of the vector meson, the physical $\rm SCET_I$ operator,
which can contribute to purely annihilation type decays, is written
by
\begin{eqnarray} Q_1=[\bar \chi_{\overline{hc}}(s\bar n) (1+\gamma_5)\gamma_\perp^\mu\eta_{\overline{hc}}+h.c. ]\bar\xi_{hc}
W_{hc}(tn)\gamma_\mu^\perp (1-\gamma_5)h_v,\end{eqnarray}
with
\begin{eqnarray} \eta_{\overline{hc}}=- {1\over in\cdot D_{\overline{hc}}
}i\not\!\!D_{{\overline{hc}}\perp}{\not\! n\over 2}\chi_{\overline{hc}}  .
\end{eqnarray}
Although this operator is suppressed relative to the leading power SCET$_{\rm I}$ four quark operators, they share the same matching coefficients, since the relevant QCD diagrams in the matching procedure are the same.  Therefore, the hard function at one loop level can be extracted from the effective Wilson coefficients in the QCD factorization approach of nonleptonic $B$ decays \cite{Beneke:2000ry}.
Similar to the nonleptonic B decays, the hard function is also convoluted with the vector meson LCDAs defined below
\begin{eqnarray}
\langle V(p,\epsilon_1)|[\bar \chi(s\bar n)\gamma_\perp^\mu\eta+h.c. ]|0\rangle
&=&f_Vm_V \varepsilon_{1\perp}^{*\mu}\int du e^{ius\bar n\cdot
p}g_\perp ^{(v)}(u)
 \nonumber \\
 \langle V(p,\epsilon_1)|[\bar
\chi(s\bar n)\gamma_\perp^\mu\gamma_5\eta+h.c. ]|0\rangle
&=&{i\over 4}f_Vm_V \epsilon_{\perp}^{\mu\nu}
\varepsilon_{1\perp\nu}^{*}\int du e^{ius\bar n\cdot p}g_\perp
^{(a)'}(u),
\end{eqnarray}
with $\epsilon^\perp_{\mu\nu}=\epsilon_{\mu\nu \bar n n}\,/\,2$. As the hard-collinear part decouples from the soft and collinear part, the factorization for the annihilation diagram also holds.  We take $B^0 \to \phi\gamma$ decay as an example, the matrix element can be factorized as
 \begin{eqnarray}
\left \langle\phi(\varepsilon_1) \gamma(\varepsilon_2) \left |
{\cal H}_{\rm eff} \right | \bar B \right \rangle|_{\rm anni.}
&=&-{G_F\over \sqrt 2}V_{tb}V_{td}^*\left(\alpha_3-{1\over
2}\alpha_{3EW}\right){f_\phi m_\phi\over E_\gamma }
\nonumber \\ &\times &
\langle\gamma(\varepsilon_2) \left |
C_{\rm FF}
\bar\xi \not\!\varepsilon^\ast_1 (1-\gamma_5)h_v \right | \bar B  \rangle ,
\end{eqnarray}
where the anti-collinear vector meson LCDA has been convoluted with the hard function, and the effective Wilson coefficients are written by \cite{Li:2003kz}
\begin{eqnarray}
\alpha_3&=&C_3+{C_4\over N_c}+C_5+{C_6\over N_c}+{\alpha_s\over 4\pi}{C_F\over
N_C}{f^\perp_V\over f_V}(C_4V_1+C_6V_2)\nonumber \\
\alpha_{3EW}&=&C_7+{C_8\over N_c}+C_9+{C_{10}\over N_c}+{\alpha_s\over 4\pi}{C_F\over
N_C}{f^\perp_V\over f_V}(C_8V_2+C_{10}V_1)
\end{eqnarray}
with the vertex correction term
\begin{eqnarray}
V_1&=&\int_0^1du T_1(u)\left[{1\over 4}g_\perp ^{(a)'}(u)-g_\perp
^{(v)}(u)\right]\nonumber \\
 V_2&=&\int_0^1du T_2(u)\left[{1\over
4}g_\perp ^{(a)'}(u)+g_\perp ^{(v)}(u)\right]\nonumber ,
\end{eqnarray}
where \cite{Beneke:2006hg}
\begin{eqnarray}
T_1(u)&=&12\ln{m_b \over \mu}-18+g(u)\nonumber \\
T_2(u)&=&-12\ln{m_b \over \mu}+6-g(\bar u) \\ \nonumber
g(u)&=&{4-6u\over \bar u}\ln u-3i\pi+\left(2{\rm Li_2}(u)-\ln^2 u+{2\ln u\over \bar u}-(3+2\pi i)\ln u-[u\to \bar u]\right).
\end{eqnarray}
The hard function $C_{FF}$ arises from matching the   weak
current $\bar u \, \gamma_{\mu \, \perp} \, (1-\gamma_5) \, b$
onto the corresponding SCET current. The remaining $B \to \gamma$ transition matrix element containing soft and collinear field can be parameterized by
\begin{eqnarray} \langle \gamma(\varepsilon_2, p)|C_{\rm FF}
\bar\xi \gamma_\mu (1-\gamma_5)h_v|\bar B_v\rangle=E_\gamma
\varepsilon^*_{2\nu}(g_\perp^{\mu\nu}F_A+i\epsilon_\perp^{\mu\nu}F_V).
\end{eqnarray}
The $B\to\gamma $ transition form factors $F_{V,A}$ also present in the $B\to \gamma \ell \nu$ decay, which has been extensively studied \cite{Lunghi:2002ju,Bosch:2003fc,Beneke:2011nf,Braun:2012kp,Wang:2016qii,Ball:2003fq,Wang:2018wfj,Beneke:2018wjp,Shen:2018abs,Liu:2020ydl,Galda:2020epp,Shen:2020hsp}. At leading power $F_A=F_V$  due
to the left-handedness of the weak interaction current and helicity-conservation of the quark-gluon
interaction in the high-energy limit, and this symmetry relation is broken by power suppressed local contributions.
At leading power both hard function and jet function have been calculated up to two-loop level and next-to-leading logarithmic resummation has been performed.  The power suppressed symmetry-breaking local contrition and symmetry conserving high-twist contribution and resolve photon contribution are also considered. Utilizing the result of  ref.\cite{Shen:2020hsp} in our calculation,
the transition amplitude of $B^0 \to \phi\gamma$ and $B_s \to \rho^0(\omega)\gamma$ is then written by
 \begin{eqnarray}
 A(B\to\phi\gamma)|_{\rm anni}
&=&-{G_F\over \sqrt 2}\lambda_t\left(\alpha_3-{1\over
2}\alpha_{3EW}\right)ef_\phi m_\phi
(F_Ag^\perp_{\mu\nu}+iF_V\epsilon^\perp_{\mu\nu})\epsilon_2^{*\mu}\epsilon_1^{*\mu}\nonumber \\
\sqrt 2A(B_s\to\rho^0\gamma)|_{\rm anni}
&=&{G_F\over \sqrt 2}\left(\lambda_u\alpha_2-{3\over
2}\lambda_t\alpha_{3EW}\right)ef_\rho m_\rho
(F_Ag^\perp_{\mu\nu}+iF_V\epsilon^\perp_{\mu\nu})\epsilon_2^{*\mu}\epsilon_1^{*\mu} \\ \nonumber
\sqrt 2A(
B_s\to\omega\gamma)|_{\rm anni}
&=&{G_F\over \sqrt 2}\left(\lambda_u\alpha_2-\lambda_t\alpha_3-{1\over
2}\lambda_t\alpha_{3EW}\right)ef_\omega m_\omega
(F_Ag^\perp_{\mu\nu}+iF_V\epsilon^\perp_{\mu\nu})\epsilon_2^{*\mu}\epsilon_1^{*\mu},
\end{eqnarray}
where
\begin{eqnarray}
\alpha_2&=&C_2+{C_1\over N_c}+{\alpha_s\over 4\pi}{C_F\over
N_C}{f^\perp_V\over f_V}C_2V_1.
\end{eqnarray}


\begin{figure}
\hspace{1.8cm}\scalebox{0.5}[0.5]{\includegraphics{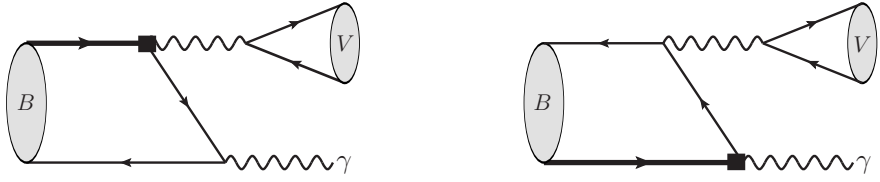}}
\caption{\label{Q7diags}Production of a vector meson via
electromagnetic penguin operator}
\end{figure}


\subsection{Contribution from electro-magnetic penguin operator}

The annihilation diagram is power suppressed because leading power four-quark operators cannot contribute to transversely polarized vector meson. However,  if the power suppressed operator $\bar \chi \gamma_\perp \eta\sim \lambda^{3/2}$ is replaced by a photon field $\not\!\!\!{\cal  A}^{em}_\perp\sim \lambda^{1/2}$, this will leads to a large enhancement factor $m_b/\Lambda$. Furthermore, for the pure annihilation type decays such as the $B^0 \to \phi \gamma$, the rather small color suppressed penguin operator Wilson coefficient will also be replaced by $C_{7\gamma}$, at the cost of an electro-magnetic coupling constant $\alpha_{em}$.
The leading order Feynman diagrams of the electro-magnetic penguin operator contribution are plotted in Fig. \ref{Q7diags}. They are corresponding to the matrix element
\begin{eqnarray}
A^{\rm LO}_{
\rm EMP}&=&  \,- {G_F\over \sqrt 2}\lambda_t \,
\int d^4 x \,  \, \langle \phi(p,\epsilon_{1}^\ast)\gamma(q,\epsilon_{2}^\ast) |
{\rm T} \left  \{ Q_qe\bar q\slashed{A} q(x),
C_{7\gamma}O_{7\gamma}(0) \right \}| \bar B(v) \rangle \nonumber \\
&& +  \left [ p \leftrightarrow q \right ] \, .
\end{eqnarray}
To evaluate this amplitude, one must have the knowledge of the matrix element of $O_{7\gamma}$. When the photon field is sandwiched between the vector meson state and
the vacuum, the matrix element reads \cite{Beneke:2005we}
\begin{eqnarray}
\langle V|e{\cal A}^{\rm em}_{\perp \mu}|0\rangle=-{2\over 3}ia_V
{e^2f_V\over m_V}\varepsilon_{\perp \mu}^*
\end{eqnarray}
with $a_\rho=3/2, a_\omega=1/2, a_\phi=-1/2$. Taking advantage of the above matrix element, the leading order result can be obtained, which has been given in \cite{Lu:2006nza}.

In this work, we make improvement by taking the QCD correction of the Wilson coefficients of $O_{7\gamma}$ operator into account. The complete ${\cal O}(\alpha_s)$ corrections including the contribution from four-quark operators and chromo-magnetic operator are accomplished in \cite{Chetyrkin:1996vx}.  Similar to the decay amplitude of $B^0 \to \gamma \gamma$ \cite{Shen:2020hfq} and $B^0 \to \gamma \ell\ell$ mode \cite{Beneke:2020fot}, the leading power contribution of
electro-magnetic dipole operator to $B^0 \to \phi \gamma$ and $B_s \to \rho^0(\omega) \gamma$ decay can be expressed as
\begin{eqnarray}
{  A}_{\rm EMP}  &=&
i \, {4 \, G_F \over \sqrt{2}}\lambda_t \, {\alpha_{\rm em} \over 4 \pi}  \left({-{2\over 3}ia_V}\right){ef_V\over m_V}\,
\varepsilon_1^{\ast \alpha}(p)  \, \varepsilon_2^{\ast \beta}(q)  \,
\left [g_{\alpha \beta}^{\perp} + i \epsilon_{\alpha \beta}^{\perp} \right ] \,
Q_q \, f_{B_q} \, m_{B_q}E_\gamma \bar m_b V_7^{eff}(0)\nonumber \\
&\times&\left[{V_7^{eff}(q^2)\over V_7^{eff}(0)}\dfrac{m_{B_{q}}}{2E_{\gamma}} \int_0^\infty{d\omega \over \omega}J(2E_\gamma,0,\omega)\phi^B_+(\omega) \right.  \nonumber \\
&&+\left. \int_0^\infty{d\omega \over \omega-{m_V^2/ m_{B_q}}}J(2E_\gamma,m_V^2,\omega)\phi^B_+(\omega)\right] .
\end{eqnarray}
The explicit expression of effective Wilson coefficient $V_7^{eff}(q^2)$ and  the jet function $J(2E_\gamma,q^2,\omega)$ at one loop level are given in \cite{Beneke:2020fot}, including the factorization scale dependence obtained from renormalization group evolution.   In our numerical analysis, the factorization scale is chosen at an intermediate scale $\mu_{hc}\sim \sqrt{ E_\gamma \omega}$ .

\subsection{Contribution from mixing of neutral vector mesons}

The mixing of the flavor-SU(3) singlet and octet states of
vector mesons to form mass eigenstates is of
fundamental importance in hadronic physics. It is commonly accepted
that the vector meson states satisfy the ``ideal" mixing,
close to the value that would lead to the complete decoupling of the light $u$ and $d$ quarks from the heavier $s$ quark
in the resultant mass eigenstates $\omega$ and $\phi$.  Actually
$\omega$ and $\phi$ are not pure states with definite
isospin  given by
\begin{equation}
\label{eq:isostates}
\state{\omega_I} =  \frac{1}{\sqrt{2}}(\state{\bar u u} +
\state{\bar
  d d})\,, \qquad
\state{\phi_I} = \state{\bar s s} \, .
\end{equation}
The mass eigenstates $\omega, \phi$ deviate from the ``ideal" states $\omega_I, \phi_I$ through a mixing matrix
 \begin{eqnarray}
\left(\begin{array}{c}|\omega\rangle \\
 |\phi\rangle\end{array}\right)= \left(
\begin{array}{cc}
\cos \delta &\sin \delta
\\-\sin \delta &\cos \delta
\end{array}\right)
\left(\begin{array}{c}|\omega_I\rangle \\
 |\phi_I\rangle
\end{array}\right) ,
\end{eqnarray}
where the mixing angle $\delta$ can be determined from the experimental data or by model calculation.
The isospin triplet $\rho^0$  can also mix with $\omega$ and $\phi$ through electro-magnetic interactions, however, the mixing angle is about one order smaller than the $\omega-\phi$ mixing, since the isospin-breaking is much smaller than the flavor SU(3) breaking effect.  Thus we do not take this isospin-breaking mixing effect into account  in our analysis.   After considering the mixing between $\omega$ and $\phi$ meson,  the $B^0 \to \phi \gamma$ and $B_s \to \omega \gamma$ decays can be expressed in terms of the decay amplitude with the ideal mixing meson final state, i.e.,
\begin{eqnarray}
A(B^0\to \phi \gamma)|_{mixing}&=& -\sin \delta A(B\to
\omega_I\gamma )\nonumber \\
A(B_s\to \omega \gamma)|_{mixing}&=& \sin \delta A(B_s\to
\phi_I\gamma ).
\end{eqnarray}

\begin{table}
\caption{Estimation of the relative size from different contributions to $B^0 \to \phi \gamma$.}\label{paraexp}
\centering
\def\arraystretch{1.2}
\setlength\tabcolsep{8pt}
\begin{tabular}{|c|c|c|c|}
\hline
Contributions  & Suppression & Enhancement & Typical value \\
\hline
${|A(B\to \phi \gamma)|_{\rm anni}\over |A(B\to \rho^0 \gamma)|_{\rm LP}}$ & ${{\alpha_3-1/2\alpha_{3EW}}\over C_{7\gamma}}\times {m_V \over m_b}$&- & 0.004\\
\hline
${|A(B\to \phi \gamma)|_{\rm EMP}\over |A(B\to \rho^0 \gamma)|_{\rm LP}}$ &$ \alpha_{\rm em}\times a_\phi $& ${m_b\over m_V}$&$0.01$ \\
\hline
${|A(B\to \phi \gamma)|_{\rm mixing}\over |A(B\to \rho^0 \gamma)|_{\rm LP}}$ & $\sin \delta$& -& 0.06\\
\hline
\end{tabular}
\end{table}

To show that the  $\omega-\phi$ meson mixing will dominate the $B^0
\to \phi \gamma$ and $B_s \to \omega \gamma$  decays,  we estimate the relative size of
different contributions to these decay modes.  For convenience,  we investigate the proportion of the absolute value of the amplitude of each contribution with respect to the absolute value of the
leading power amplitude in $B \to \rho^0 \gamma$ decay in Table~\ref{paraexp}.
From this table we can see that the mixing effect  can increase
the branching ratio from annihilation topology in QCD factorization approach over two orders of magnitudes, and is   one order larger than the contribution from electro-magnetic operators. For $B_s \to \rho^0(\omega) \gamma$ decays, the tree operators with large Wilson coefficient can contribute, while they are suppressed by a small suppression factor from CKM matrix elements, i.e., $|V_{ub}V_{us}/V_{tb}V_{ts}|$, therefore, the relative size of different type of contribution in $B_s \to \rho^0(\omega) \gamma$ decays is similar to $B \to \phi\gamma$.

\section{Numerical analysis}
\subsection{ Input parameters}
The decay amplitudes for the $B^0 \to \phi\gamma$ and $B_s \to \rho^0(\omega)\gamma$ decays have been obtained in the previous section, they will be utilized to predict the branching ratios of these decay modes. Firstly we specify the input parameters which will be used in the numerical calculation. Among various parameters, the mixing angle $\delta$ is of unique importance because it will provide the major source of uncertainties in our calculations. The mixing angle has been discussed in many phenomenological methods such as  the framework of the hidden local symmetry Lagrangian \cite{Benayoun:2007cu,Benayoun:2009im}, the chiral perturbation theory \cite{Klingl:1996by,Kucukarslan:2006wk}, the light front quark model \cite{Choi:2015ywa} and the Nambu-Jona-Lasinio model \cite{Vogl:1991qt,Klevansky:1992qe} etc., with the obtained values varying at the interval about $3^\circ \sim 5^\circ$  (most of the studies prefer $[3^\circ ,4^\circ]$). In this work, we adopt the value of mixing angle as $\delta=3.5^\circ\pm 0.5^\circ$.

To arrive at the result of the decay amplitudes from the $\omega-\phi$ mixing, the leading power contribution of $B^0\to \omega \gamma$ and $B_s \to \phi \gamma$ is necessary.  The basic nonperturbative inputs in these amplitudes are the soft form factor $\zeta^{\rm BV}_\perp$ and the light-cone distribution amplitude of $B$-meson and $\rho,\omega,\phi$ meson.  The soft factors $\zeta^{\rm BV}_\perp$   defined in terms of the matrix element of SCET$_{\rm I}$ operators have been calculated using SCET sum rules. The complete NLO corrections to the correlation function as well as the power suppressed higher twist contribution have been calculated in ref.\cite{Gao:2019lta}. The result is adopted as
 $\zeta^{\rm BV}_\perp=0.33\pm0.10$.  For the $B_s \to \phi$ transition, the result is  $\zeta^{\rm B_sV}_\perp=0.35\pm0.10$, allowing a small SU(3) breaking effect.

For the leading twist two-particle  $B$-meson distribution amplitude, we will employ the following three-parameter model
\begin{eqnarray}
\phi_B^+(\omega)={\Gamma(\beta)\over \Gamma(\alpha)}\,{\omega\over \omega^2_0}\,e^{-{\omega\over \omega_0}}\,U\left(\beta-\alpha,3-\alpha,{\omega\over \omega_0}\right),\end{eqnarray}
where $U(\alpha,\gamma,x)$ is the confluent hypergeometric function of the second kind.
 A special case is the exponential model when $\alpha=\beta$
 \begin{eqnarray}
\phi^+_B(\omega)={\omega\over \omega^2_0}\,e^{-{\omega\over \omega_0}}.
\end{eqnarray}
To estimate the error from the models, we will let $\alpha-\beta$ vary at the region $-0.5<\alpha-\beta<0.5$, then we employ two models with
 $\alpha=2.0, \beta=1.5$ and $\alpha=1.5, \beta=2.0$. The parameter $\omega_0$ is closely related to the first inverse moment $1/\lambda_B$,  whose determination
has been discussed extensively in the context of exclusive $B$-meson decays
(see \cite{Wang:2015vgv,Wang:2017jow,Lu:2018cfc,Gao:2019lta} for more discussions). Here we will employ
$\lambda_B(1 \, {\rm GeV})=0.35\pm0.05 \, {\rm GeV}$  and  $\lambda_{B_s}(1 \, {\rm GeV})=0.40\pm0.05 \, {\rm GeV}$. For the light vector meson, the leading twist LCDAs can be expanded in terms of Gegenbauer polynomials due to the behavior of scale evolution, i.e.
\begin{eqnarray}
\phi_V(x)&=&6x(1-x)[1+a_VC_2^{3/2}(2x-1)]\nonumber\\
\phi_{V_\perp}(x)&=&6x(1-x)[1+a_{V_\perp}C_2^{3/2}(2x-1)].
\end{eqnarray}
For the power suppressed vector meson LCDAs, ignoring the three parton wave function, we   have  the following expression \cite{Cheng:2008gxa}
\begin{eqnarray}
{1\over 4}g_\perp ^{(a)'}(u)-g_\perp
^{(v)}(u)=-\int_0^u{\phi_V(v)\over \bar v},\,\,\,\,\,{1\over 4}g_\perp ^{(a)'}(u)+g_\perp
^{(v)}(u)=\int_u^1{\phi_V(v)\over  v}.
\end{eqnarray}
In the annihilation topology, the $B_{(s)}\to \gamma$ transition form factors $F_{V,A}$ is required. The SU(3) breaking effect is found to be negligible after taking the next-to-leading power contribution into account, thus we adopt the same result for both $B$ and $B_s$ decays, i.e., $F_V=0.23\pm0.07$, $F_A=0.21\pm 0.07$.  The values of the other parameters are presented in Table \ref{table:para}.

\begin{table}
	\caption{Input parameters}  \label{table:para} 	
	\centering
	\def\arraystretch{1.5}
	\setlength\tabcolsep{8pt}
	\begin{tabular}{c c|c c}
		\hline
		\hline
		$\tau_{B^{0}}$ & 1.52ps & $G_{F}$ & $1.116637\times10^{-5}$\\
		$\tau_{B_s}$ & 1.51ps & $\lambda$ & 0.22650 \\
		$f_{B}$ & 0.192 &  $\bar{\rho}$ & 0.141\\
		$f_{B_s}$ & 0.230 & $A$ & 0.790 \\
		$\sigma^{(1)}_{B}$ & $1.63\pm 0.15$ & $\bar{\eta}$ & 0.357 \\
		$\sigma^{(1)}_{B_s}$ & $1.49\pm 0.15$ & \\
		\hline
		$f_{\rho}$(1GeV) & 0.216$\pm$ 0.003 & $a_{2\rho}$(1GeV) & 0.15$\pm$0.07  \\
		$f_{\omega}$(1GeV) & 0.187$\pm$0.005 & $a_{2\omega}$(1GeV) & 0.15$\pm$0.07 \\		
		$f_{\phi}$(1GeV) & 0.215$\pm$0.005 & $a_{2\phi}$(1GeV) & 0.18$\pm$0.08\\
		$f_{\rho\perp}$(1GeV) & 0.165$\pm$ 0.009 & $a_{2\rho\perp}$(1GeV) & 0.14$\pm$0.06  \\		
		$f_{\omega\perp}$(1GeV) & 0.151$\pm$0.009 & $a_{2\omega\perp}$(1GeV) & 0.14$\pm$0.06 \\		
		$f_{\phi\perp}$(1GeV) & 0.186$\pm$0.009 & $a_{2\phi\perp}$(1GeV) & 0.14$\pm$0.07\\	
		\hline	
		\hline
	\end{tabular}
\end{table}

\subsection{Phenomenological predictions}

Collecting all the contributions to the factorization amplitudes calculation in the previous section together, we arrive at the final expression of the decay amplitudes for the pure annihilation type $B \to V \gamma$ decays
\begin{eqnarray}
\begin{aligned}
&A(B^0\to \phi\gamma)=\cos \delta[A(B^0\to \phi_I\gamma)|_{\rm anni}+A(B^0\to \phi_I\gamma)|_{\rm
EMP}]-\sin \delta A(B\to\omega_I\gamma)\\
&A(B_s\to \omega \gamma)=\cos \delta[A(B_s\to \omega_I \gamma)|_{\rm anni}+A(B_s\to \omega_I \gamma)|_{\rm
EMP}]+\sin \delta A(B_s\to\phi_I\gamma)\\
&A(B_s\to \rho^0 \gamma)= A(B_s\to \rho^0 \gamma)|_{\rm anni}+A(B_s\to \rho^0 \gamma)|_{\rm
EMP} .
\end{aligned}
\end{eqnarray}
The results for the phenomenological observables
in pure annihilation type  decays are then studied.
As the decay rates are relatively small and the observables such as CP asymmetry are hard to be detected, we concentrate on the CP-averaged branching ratios defined below
\begin{eqnarray}
\langle B(B^0\to V\gamma)\rangle={B(\bar B^0 \to V\gamma)+B(B^0 \to V\gamma)\over 2},
\end{eqnarray}
where the specific expression of the branching ratio is give by
\begin{equation}
	B(\bar B^0 \to V\gamma)=\dfrac{\tau_{B}}{16\pi m_B}\bigg(1-\dfrac{m_{V}^{2}}{m_{B}^{2}}\bigg)|A(\bar B^0\rightarrow V\gamma)|^{2} .
\end{equation}
To illustrate the contribution from various sources, we firstly present the results of each kind of contributions in Table \ref{contributions}. In the contribution from mixing of neutral vector mesons, we only consider the leading power contribution to the $B^0 \to \omega_I\gamma$ and $B_s\to \phi_I\gamma$ amplitudes, because the mixing angle is already a small quantity. The QCD factorization result of the pure annihilation contribution is consistent with the result in \cite{Li:2003kz}.  The contribution from elector-magnetic penguin operator is a bit larger than our previous predictions in \cite{Lu:2006nza}  as  the  leading logarithm resummation of effective Wilson coefficient $C_{7eff}$ is employed and some parameters are renewed. Our results indicate that the  branching ratio of $B^0 \to \phi\gamma$ purely from the $\phi-\omega$ mixing is three orders larger than that from the annihilation topology, and also about two orders larger than that from the electro-magnetic penguin contribution in the decays. Apparently this result is consistent with our rough estimation.  Taking advantage of the central values  in Table~\ref{paraexp}, the total branching ratio of $B^0 \to \phi \gamma$ is obtained as $3.99\times 10^{-9}$, which have the chance to be measured in the Belle-II with  an ultimate integrated luminosity of $50 ab^{-1}$. The $B_s \to \omega \gamma$ decay with the branching ratio $2.01\times 10^{-7}$ can also confront the LHC-b data.

\begin{table}
	\caption{Branching fractions of different contributions.}\label{contributions}
	\centering
	\def\arraystretch{1.2}
	\setlength\tabcolsep{8pt}
	\begin{tabular}{c|c|c|c|c}
		\hline
		\hline
		Channels   & $A_{anni}$ Only  & $A_{EMP}$ Only & $A_{mixing}$ Only  & Total \\
		\hline
		$B(B^{0}\rightarrow\phi\gamma)$ & $3.93\times10^{-12}$ & $2.02\times10^{-11}$&$3.61\times10^{-9}$&$3.99\times10^{-9}$\\
		\hline
		$B(B_s^{0}\rightarrow\omega^{0}\gamma)$ & $6.94\times10^{-11}$ & $3.91\times10^{-10}$&$1.80\times10^{-7}$&$2.01\times10^{-7}$\\
		\hline
		$B(B_s^{0}\rightarrow\rho^{0}\gamma)$ & $4.89\times10^{-11}$ & $4.79\times10^{-9}$&-&$5.67\times10^{-9}$\\
		\hline
		\hline
	\end{tabular}
\end{table}

We define the following ratios of branching fractions, which can highlight the importance of the vector meson mixing effects
\begin{eqnarray}
R_{\rho\omega}={B(B_s^{0}\rightarrow\rho^{0}\gamma)\over B(B_s^{0}\rightarrow\omega^{0}\gamma)},\,\,\,R_{\rho\phi}={B(B_s^{0}\rightarrow\rho^{0}\gamma)\over B(B^{0}\rightarrow\phi\gamma)}.
\end{eqnarray}
Naively considering, the first ratio $R_{\rho\omega}\sim 1$, since the Feynman diagrams of the annihilation topology are the same for $B_s \to \rho^0\gamma$ and $\omega\gamma$ decays, furthermore, the contribution from  electro-magnetic penguin will even enhance this ratio to 10, as $|a_\rho/a_\omega|=3$.  The second ratio $R_{\rho\phi}$ is expected to be large for the CKM enhancement from the ratio  $|V_{tb}V_{ts}/V_{tb}V_{td}|^2$. While after the $\phi-\omega$ mixing effect is taken into account,  the values of these ratios are dramatically changed. Our result shows that $R_{\rho\omega}\simeq 0.03$ which confirms the dominance of meson mixing effects, and $R_{\rho\phi}\simeq 1.4$ which indicates that the contribution from $\omega-\phi$ mixing is larger than the electro-magnetic penguin amplitude by a factor  $|V_{tb}V_{ts}/V_{tb}V_{td}|$ approximately. The predicted values of these ratios are expected to be tested in the future experiments.

Now we investigate the theoretical uncertainties.  Inspecting the distinct sources of the yielding theory uncertainties as collected  in the following formula, we have
\begin{eqnarray}
B(B^{0}\rightarrow\phi\gamma)&=&3.99\;^{+1.19}_{-1.04}|_{\zeta_{\perp}}\;^{+1.16}_{-1.01}|_{\delta}\;^{+0.02}_{-0.02}|_{f_{V},\phi_V}\;^{+0.08}_{-0.06}|_{\lambda_B}\;^{+0.07}_{-0.15}|_{\phi_B}\;^{+0.01}_{-0.01}|_{F_{V}}\;^{+0.01}_{-0.01}|_{F_{A}}\times10^{-9},\nonumber \\
B(B_s\rightarrow\omega\gamma)&=&2.01\;^{+0.56}_{-0.49}|_{\zeta_{\perp}}\;^{+0.58}_{-0.51}|_{\delta}\;^{+0.01}_{-0.01}|_{f_{V},\phi_V}\;^{+0.02}_{-0.02}|_{\lambda_B}\;^{+0.00}_{-0.02}|_{\phi_B}\;^{+0.01}_{-0.01}|_{F_{V}}\;^{+0.01}_{-0.01}|_{F_{A}}\times10^{-7},\nonumber  \\
B(B_{s}\rightarrow\rho^0\gamma)&=&5.67\;^{+0.16}_{-0.16}|_{f_{V},\phi_V}\;^{+0.55}_{-0.41}|_{\lambda_B}\;^{+1.50}_{-1.69}|_{\phi_{B}}\;^{+0.04}_{-0.04}|_{\sigma_B}\;^{+0.15}_{-0.15}|_{F_{V}}\;^{+0.15}_{-0.15}|_{F_{A}}\times10^{-9} .
\end{eqnarray}
It is obvious that  the soft form factors which play the dominant role in the $B \to \omega \gamma$ and $B_s \to \phi \gamma$ decays provides an important source of uncertainties. The mixing angle between $\phi$ and $\omega$ meson is another major source of uncertainty as expected. As the decay amplitudes of $B \to \phi \gamma$ and $B_s \to \omega \gamma$ are very sensitive to  the  vector meson mixing effect, these channels can serve as a good platform to determine the mixing angle, i.e., the mixing angle between $\omega$ and $\phi$ meson can be determined by
\begin{eqnarray}
\sin \delta \simeq\sqrt{B(B^{0}\rightarrow\phi\gamma)\over B(B^{0}\rightarrow\omega\gamma)}
\,\,\, or \,\,\,
\sqrt{B(B_s^{0}\rightarrow\omega\gamma)\over B(B_s^{0}\rightarrow\phi\gamma)},
\end{eqnarray}
if the related decay modes are measured. For the $B(B_{s}\rightarrow\rho^0\gamma)$ decay which is dominated by the electro-magnetic penguin operator, the major source of uncertainty is from the shape and the first inverse moment of the LCDA of $B_s$ meson. Therefore, it is of great importance to improve the study of  $B_{(s)}$ meson LCDA. A recent effort is the introduction of the quasi-parton distribution amplitude of $B$ meson \cite{Wang:2019msf} so that it can be calculated by lattice QCD simulation.

\section{Closing remarks}

The pure annihilation type radiative B meson decays, including $B^0 \to \phi \gamma$ and $B_s \to \rho^0(\omega)\gamma$ decays, are very rare in the standard model, which makes them very sensitive to the new physics signals beyond the standard model. We reviewed factorization of $B \to V\gamma$ decays at leading power using SCET, and derived the factorization formula for annihilation topology. The electro-magnetic penguin contribution to the pure annihilation radiative decays, which is power enhanced, is also revisited with leading logarithm resummation of the effective Wilson coefficients taken into account. As the major subject of this work, we studied the contribution of the neutral vector meson $\omega-\phi$ mixing to the decay amplitudes. Although the mixing angle of the $\phi-\omega$ is only a few percent, this contribution owns larger Wilson coefficients as well as power enhancement compared with annihilation topology. A rough estimate indicates that the contribution from  $\phi-\omega$ mixing is dominant in the pure annihilation radiative decays.  The  numerical calculation shows that the  branching ratio of $B^0 \to \phi\gamma$ purely from the $\phi-\omega$ mixing is three orders larger than that from the annihilation topology, and also two orders larger than that from the electro-magnetic penguin contribution in the decays.  The similar hierarchy between the different contributions holds for $B^0 \to \omega\gamma$. The decay rate of $B_s \to \rho^0\gamma$ is much smaller than that of $B_s \to \omega\gamma$ for the suppressed mixing effect is not considered. The new defined ratios $R_{\rho\omega}\simeq 0.03$ and $R_{\rho\phi}\simeq 1.4$ further highlight the importance of the mixing effect. The predicted branching ratios of  $B^0 \to \phi \gamma$ and $B_s \to \rho^0(\omega)\gamma$ decays are given below:
\begin{eqnarray}
B(B^{0}\rightarrow\phi\gamma)&=&3.99^{+1.67}_{-1.46} \times10^{-9},\nonumber \\
B(B_s\rightarrow\omega\gamma)&=&2.01^{+0.81}_{-0.71} \times10^{-7}, \\
B(B_{s}\rightarrow\rho^0\gamma)&=&5.67^{+1.62}_{-1.76} \times10^{-9}.\nonumber
\end{eqnarray}
These results are to be tested by the Belle-II and LHC-b experiments.

\section*{Acknowledgement}
This work is partly supported by the national science foundation of China under contract number 11521505 and 12070131001, and by the science foundation of Shandong province under contract number ZR2020MA093.

\end{document}